\documentclass{jkas}

\def\year{2017} 
\def\month{June} 
\def\volume{50} 
\def\issue{3} 
\def\beginpage{71} 
\def\endpage{78} 
\def\received{April 18, 2017} 
\def\accepted{June 7, 2017} 
\date{Received \received; accepted \accepted}

\usepackage{flushend} 
\usepackage{microtype}

\title{
Seoul National University Camera II (SNUCAM-II): The New\\ SED Camera for the Lee Sang Gak Telescope (LSGT)
}

\author[]{Changsu Choi}
\author[]{Myungshin Im}

\affil[]{Center for the Exploration of the Origin of the Universe, Department of Physics and Astronomy, Seoul National University, Gwanak-gu, Seoul 08826, Korea; \email{changsu@astro.snu.ac.kr, mim@astro.snu.ac.kr}}

\def\corrauthor{
M. Im
}

\def\runningauthor{
Choi \& Im
}

\def\runningtitle{
SNUCAM-II: The New SED Camera for the LSGT
}

\def\keywords{
instrumentation: CCD camera, medium-band filters
}

\def\abstracttext{
We present the characteristics and the performance of the new CCD camera system, SNUCAM-II (Seoul National University CAMera system II) that was installed on the Lee Sang Gak Telescope (LSGT) at the Siding Spring Observatory in 2016. SNUCAM-II consists of a deep depletion chip covering a wide wavelength from 0.3 $\mu$m to 1.1 $\mu$m with high sensitivity (QE at $> 80$\% over 0.4 to 0.9 $\mu$m). It is equipped with the SDSS $ugriz$ filters and 13 medium band width (50 nm) filters, enabling us to study spectral energy distributions (SEDs) of diverse objects from extragalactic sources to solar system objects. On LSGT, SNUCAM-II offers 15.7 $\times$ 15.7 arcmin field-of-view (FOV) at a pixel scale of 0.92 arcsec and a limiting magnitude of $g$ = 19.91 AB mag and z=18.20 AB mag at 5$\sigma$ with 180 sec exposure time for point source detection.
}

\begin{document}
\jkashead 


\section{Introduction\label{sec:introduction}}

Even in the era of giant telescopes, small telescopes -- telescopes with aperture size $\lesssim 0.5$\,m -- are still valuable observational facilities. They can cover a large field of view thanks to short focal lengths, and they can be dedicated for a few specific research projects due to the availability of the telescope time. Furthermore, small telescopes can be relatively easily deployed for robotic operations and save observation time for astronomers. Therefore, monitoring observations and rapid follow-up observations of objects like active galactic nuclei (AGNs), Gamma Ray Bursts (GRBs) and supernovae (SNe) are routinely carried out with small telescopes (Drake et al. 2009; Klotz et al. 2008; Shappee et al. 2014).
The major disadvantage of small telescopes is the lack of light gathering power. Currently, the best we can do to augment the lack of light gathering power of a single small telescope is to use CCD cameras with high Quantum Efficiency (QE). Yet, many small telescopes are equipped with commercially available font-illuminated CCD cameras that have very low QEs at short ($\sim 450$ nm) and long ($> 800$ nm) wavelengths.

Recently, we began operating a robotic 0.43\,m telescope, the Lee Sang Gak Telescope (LSGT; Im et al. 2015a), and carrying out a monitoring survey of nearby galaxies with it. To maximize the observing efficiency of LSGT, we have assembled a new CCD camera system, Seoul National University CCD Camera II (SNUCAM-II), by utilizing a commercially available CCD camera with a deep depletion chip that has QEs at around 80\% over 400 -- 900 nm, and a 20 position filter wheel that can house 13 medium-band filters and five broad-band filters. The camera's QE represents improvement in QE by factors of 1.5 to 3 at 400 nm and 3 to 20 at 900 nm, in comparison to commonly available CCDs such as KAI-16000 and KAF-3200ME (see Figure 3 in Im et al. 2015a) making this telescope as one of the most sensitive small telescopes in the world in its class. The camera is named as a successor of SNUCAM (Im et al. 2010), a sensitive 4k$\times$4k CCD camera, and SNUCAM-II improves upon SNUCAM in sensitivity at long wavelengths out to $Y$-band regime, making it possible for this instrument to study high redshift quasars and GRBs (e.g., Choi et al. 2012).

Another unique aspect of SNUCAM-II is the availability of many medium-band filters. The use of medium-band filters makes it possible to trace the SEDs of objects of interest, and this is especially advantageous for monitoring observation of AGNs where the time lag of the broad emission lines and the continuum flux can be traced efficiently for black hole mass measurement. We have demonstrated the power of the medium-band observation on a moderate-sized telescope using the SED Camera for Quasars in Early Universe (SQUEAN; Kim et al. 2016) on the 2.1\,m telescope at the McDonald observatory where 9 medium-band filters are installed. Specifically, we have shown that medium-band observation can select high redshift quasars effectively (Jeon et al. 2016), and we have been using the instrument to survey faint quasars at $z \sim 5$. Many other applications of the medium-band imaging are possible by tracing SEDs of various objects such as asteroids, stars, galaxies, to name a few.
In this paper we describe the overall characteristics of the SNUCAM-II system, and its performance that are derived from the laboratory testing and the on-sky observation using LSGT.

\section{System}

Here, we describe each component of SNUCAM-II. Those are CCD camera, band pass filters and a filter wheel. Software to control LSGT system and typical procedure of remote observation are explained here to help the potential observer.

\begin{figure}[t!]
\centering
\includegraphics[width=84mm]{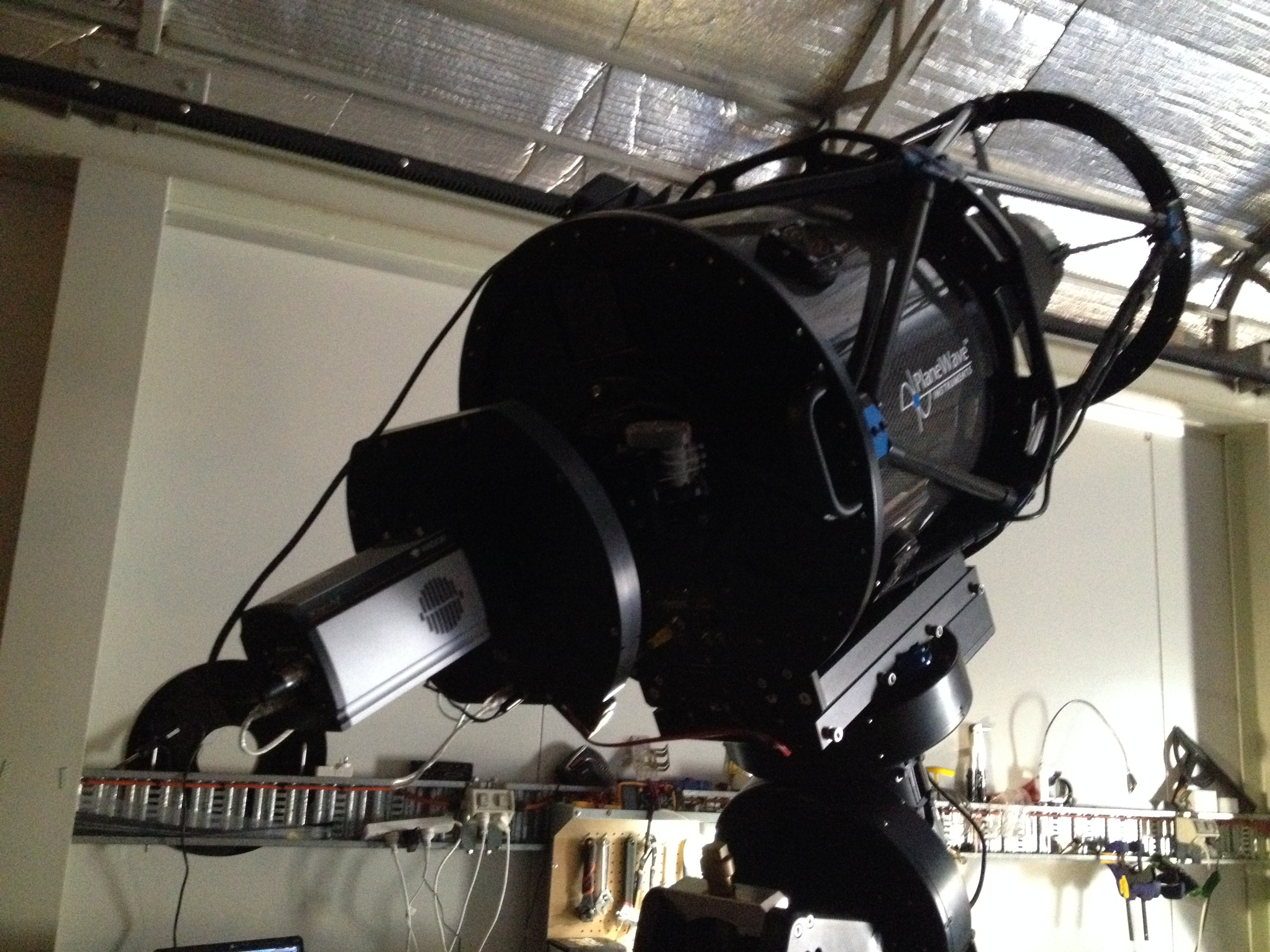}
\caption{SNUCAM-II after installation at LSGT.}
\end{figure}

\subsection{CCD Camera}

Due to a wide variety of the applications we expect for SNUCAM-II, we employed a CCD camera with good sensitivity over a broad wavelength region from 300 nm to 1100 nm. We adopted the ANDOR\footnote{\url{http://www.andor.com}} iKon-M DU934P BEX2-DD model CCD camera that uses a deep depletion, back illuminated and fringe suppression E2V CCD chip with extended range dual Anti Reflection coating on it. Its QE is about 90\% from 400 nm to 800 nm and even at longer wavelength 1000nm, QE is over 30\% thanks to the deep depletion chip (the black dashed line in Figure 2). Each pixel has a full well depth of 130,000 $e^{-}$. The CCD chip has 1024$\times$1024 pixels and the physical size of each pixel is 13 $\mu$m $\times$ 13 $\mu$m, which translates into 0.92 arcsec pixel scale at the $f/6.8$ focal plane of LSGT. We use a model with thermoelectric cooling that can cool the CCD chip to $-80^{\circ}$C. For our nominal operation, we set its cooling temperature as $-70^{\circ}$C to achieve a reasonable balance between the QE at long wavelength and the reduction in dark current (see Park et al. 2012). The camera offers 4 different readout rates (5.0, 3.0, 1.0, 0.05 MHz) as a default, and users can choose parameters of Vertical Shift Speed (VSS), Horizontal Shift Speed (HSS) and Preamp Gain (PG). VSS and HSS can be set to one of 2.25, 4.25, 8.25, 16.25, 32.25, 64.25 $\mu$sec. PG has three kinds of values as $\times$1, $\times$2 and $\times$4. Also binning can be configured as one of following setting, 1$\times$1, 2$\times$2, 4$\times$4, 8$\times$8 and 16$\times$16. Shutter speed is at the default value of 30 msec at opening and closing and adjustable. For mechanical connection to the filter wheel, C mount is installed on top of CCD camera, so we used a custom made adapter for connecting CCD camera to the filter wheel. As for the setting of the parameters, we use the following values as default: PG of $\times$4 (gain $\sim$ 1.15 $\pm$ 0.03 $e^{-}$/ADU), 1 MHz readout rate, and the cooling temperature of $-70^{\circ}$C. VSS, HSS, and the shutter speed are set at 4.25 $\mu$sec, 2.25 $\mu$sec and 30 msec respectively as recommended by the manufacturer.

\begin{figure}[t!]
\centering
\includegraphics[trim=8mm 2mm 12mm 13mm, clip, width=84mm]{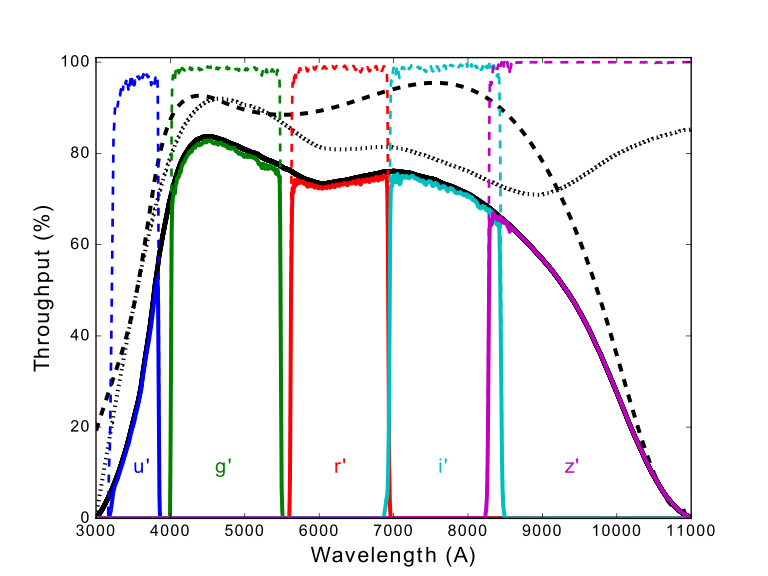}
\caption{Transmission curves of the SDSS $ugriz$ filters (colored dashed/solid line) and total throughput of the SNUCAM-II system (thick solid solid line), taking into account the CCD QE (black dashed line) and the throughput of the telescope optics (black dotted line; Im et al. 2015a). The colored solid lines show the filter transmission curves after taking into account the CCD QE and the optics throughput.}
\end{figure}

\subsection{Filters and Filter Wheel}

The SNUCAM-II uses a default set of 18 circular filters with 25 mm diameter each: SDSS $ugriz$ filters and 13 medium band filters. The filter transmission curves and the transmission curves multiplied by the CCD QE and the throughput of the telescope optics are presented in Figures 2 and 3. Table 1 shows the filter names, their effective wavelengths and FWHM (Full Width at Half Maximum) values. The SDSS $ugriz$ filters are purchased from the Astrodon company (their Generation 2 Sloan filters)\footnote{\url{http://www.astrodon.com}}. The 13 medium band width (50 nm) filters spanning from 400nm to 1050nm are standard products from Edmund Optics.\footnote{\url{http://www.edmundoptics.co.kr}} We named them by adding the central wavelength in units of nm to the initial $m$ meaning `medium band filter' (e.g., m425 for the medium band filter with the central wavelength at 425 nm). These 18 filters are mounted on the Finger Lake Instrumentation LLC\footnote{\url{http://www.flicamera.com}} (FLI) CenterLine-1-20 color filter wheel. The filter wheel has dual filter wheels Wheel 0 and Wheel 1, with 10 slots each. Keeping the first slot of each wheel as blank position, 18 filters can be installed. After installation of the filters, the physical positions of the filters and logical filter names used in programs are connected by naming in `FLIfilters' program provided by the manufacturer.

\begin{figure}[t!]
\centering
\includegraphics[trim=8mm 0mm 12mm 13mm, clip, width=84mm]{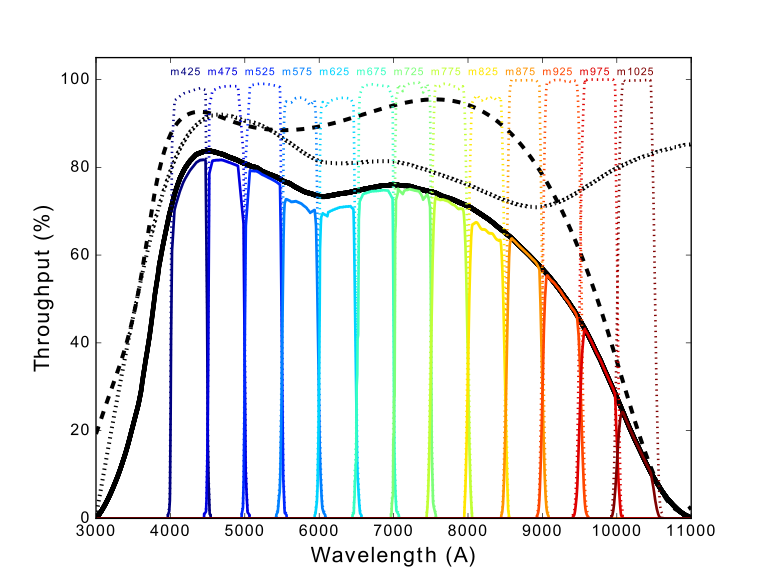}
\caption{Transmission curves of the medium-band filters before (colored dotted line) and after (colored solid lines) taking into account the CCD QE and the telescope optics throughput. The meanings of the other lines are identical to those in Figure 2.}
\end{figure}

\subsection{Software and Data Collection}

For the camera system control, we use commercial softwares: `MaximDL pro 5'\footnote{\url{http://diffractionlimited.com}} and `FLIfilter' programs. These programs and the telescope operating system software are integrated into the ACP Observatory Control Software for the command tasks necessary for the observations (observation program reservation, remote communication, actual performance of the system and the roof control). A web page is set up for system status monitoring, the observation plan, the observation program reservation and the weather information of SSO. An observer can configure all the parameters of the actual observation on the webpage and monitor activities of LSGT and SNUCAM-II in real time. A common mode of the observation uses the `plan' document that defines the target coordinate, the number of images to take, the filter, and the binning and the exposure times. This `plan' document is uploaded to the web-based reservation system for robotic observation of the target. When the observation plan is finished, the system sends an observation summary e-mail to observer along with the weather data figure, logs and the list of the obtained data. The images are uploaded on the data storage server where an observer can access with ID and password and download them. Currently, these systems run on the Window 7 operating system. The schematic diagram of the system is shown in Figure 4. The telescope and the computer system is under management of iTelescope.net.

\begin{figure}[t!]
\centering
\includegraphics[trim=8mm 5.5mm 10mm 5mm, clip, width=84mm]{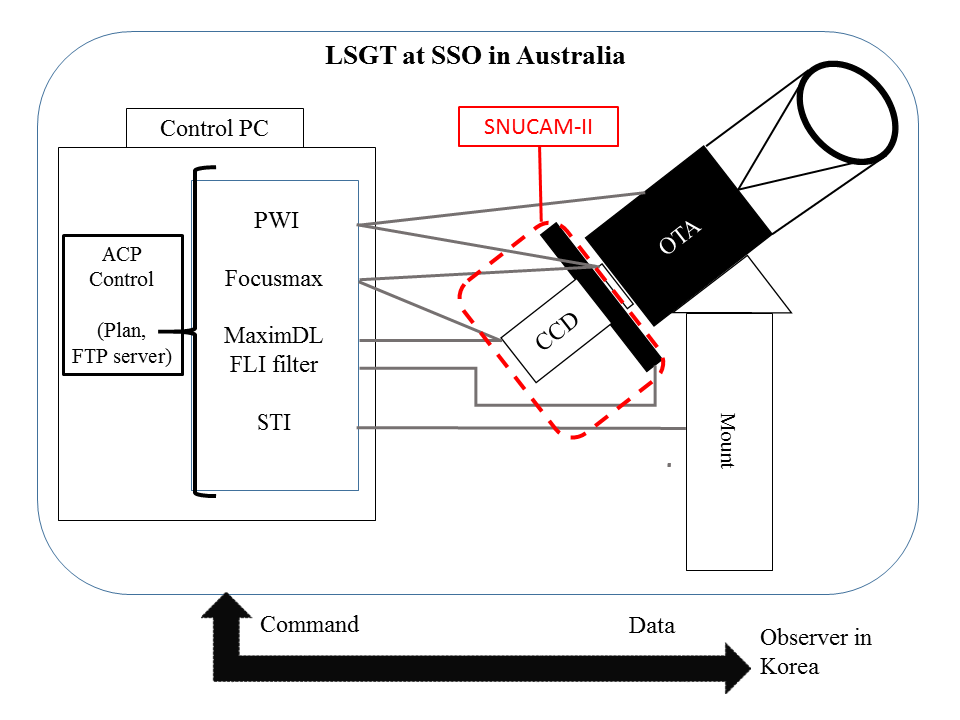}
\caption{Schematic diagram of the SNUCAM-II system.}
\end{figure}

\begin{table}[t!]
\caption{Filters of SNUCAM-II  \label{tab:snucamtable1}}
\centering
\begin{tabular}{lrr}
\toprule
Filter & Effective & FWHM  \\
Name   & Wavelength & (nm) \\
       & (nm)       &      \\
\midrule
$u$     &  363.1  &  25.7 \\
$g$ 	&  474.5  & 147.5 \\
$r$ 	&  627.7  & 131.4 \\
$i$ 	&  767.7  & 149.8 \\
$z$ 	&  915.4  & 157.2 \\
$m425$ 	&  425.9  &  48.5 \\
$m475$ 	&  475.0  &  48.6 \\
$m525$ 	&  524.9  &  49.7 \\
$m575$ 	&  574.3  &  48.5 \\
$m625$ 	&  624.7  &  49.0 \\
$m675$ 	&  675.7  &  48.6 \\
$m725$ 	&  725.7  &  48.9 \\
$m775$	&  774.8  &  49.4 \\
$m825$ 	&  824.5  &  49.3 \\
$m875$ 	&  875.4  &  48.2 \\
$m925$	&  925.1  &  51.8 \\
$m975$ 	&  973.4  &  49.1 \\
$m1025$ & 1021.0  &  40.6 \\
\bottomrule
\end{tabular}
\tabnote{
A list of the currently available SNUCAM-II filters. The effective wavelengths and the FWHM values of 18 filters are described together. The effective wavelengths are derived following equation (3) of Fukugita et al. (1996). The FWHM values are widths between 50\% of peak values of transmission curves.
}
\end{table}

\section{Characteristics of SNUCAM-II\label{sec:characteristic}}

In this section, we characterize the properties of SNUCAM-II camera. The characteristics of SNUCAM-II are summarized in Table 2.

\subsection{Dark Current}

We took bias and flat images in the laboratory and on the sky. By combining the bias frame values and the flat images taken at two different epochs ($B_{1}$ and $B_{2}$ for bias, $F_{1}$ and $F_{2}$ for flats), we calculated the gain and the readout noise values using the equations below (Howell et al. 2006).
\begin{equation}
Gain = \frac{(\bar{F}_{1} + \bar{F}_{2}) - (\bar{B}_{1} + \bar{B}_{2})}{\sigma^{2}_{F_{1} - F_{2}} - \sigma^{2}_{B_{1} - B_{2}}}
\end{equation}
\begin{equation}
Read\ Noise = \frac{Gain \cdot \sigma_{B_{1} - B_{2}}}{\sqrt{2}}
\end{equation}
We find that the gain and the readout noise are 1.15 $\pm$ 0.03 $e^{-}$/ADU (Analog-to-Digital Unit) and 6.0 $\pm$ 0.1 $e^{-}$, consistent with the test report values from the manufacturer.

\subsection{Bias, Dark, and Flat}

The dark current values were measured in the laboratory and in the dome during the daytime in dark condition with the dome light turned off. The temperature settings were varied from $-80^{\circ}$C to $-30^{\circ}$C, and the mean values of the frames from 300 sec exposures were recorded after bias subtraction. The result is shown in Figure 5, where the blue circles are measured values and the green circles are values from test report of the manufacturer. The laboratory values are consistent with the test report, with the dark current of 0.2 e$^-$/s/pixel at $-70^{\circ}$C setting. We checked the variability of the bias image, dark current and pixel-to-pixel variation of flat images. Compared to the readout noise of 6.6 $e^{-}$ of the optimal setting from the test report, the measured value from multiple test bias data is 6.0 $e^{-}$, consistent with the test report results. And its variation over different locations in a single image is up to 0.5\%. There is a small amount of intra-night variability which may originate from electrical instability. We examined the bias variation over several hours and found that the mean bias fluctuation level is lower than 4 ADUs which is less than the readout noise value but not negligible. We also compared the flat images taken at different nights, and examined the pixel-to-pixel variation in normalized flat images. The peak-to-peak variation of flat images is less than 5\% for all the filters. These calibration frames are taken regularly every month. So far, there is no noticeable variation in these frames for 6 months since the regular use of SNUCAM-II started.

\begin{figure}[t!]
\centering
\includegraphics[trim=8mm 1mm 17mm 7mm, clip, width=84mm]{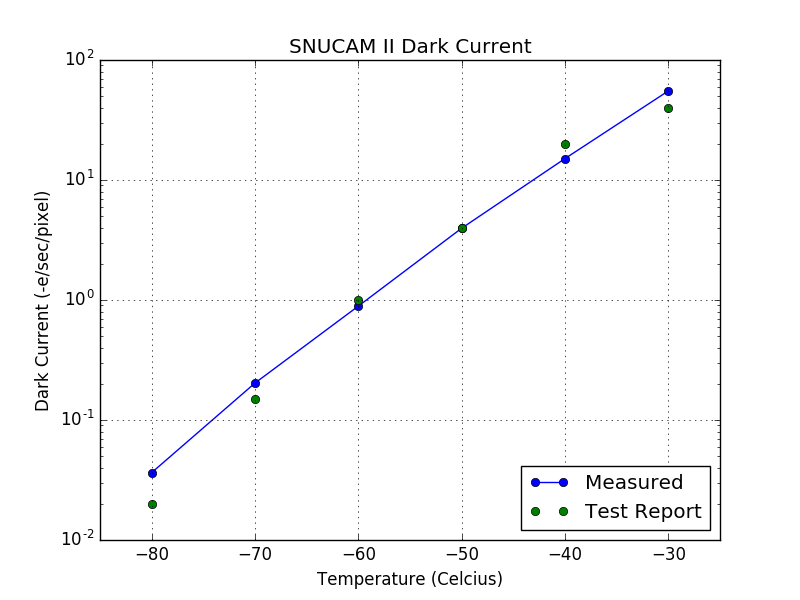}
\caption{Dark current of SNUCAM-II. Note that the dark current is very small at $-70^{\circ}$C. }
\end{figure}

\subsection{Linearity}

We examined the linearity of the detector in the laboratory. In a dark room, a light source is placed to create a diffuse background light. Then, the mean values of the image are recorded by varying the exposure time. The result is plotted in Figure 6. The mean values are well fitted with a linear relation with a constant slope up to 60000 ADU level. Even at a low signal level of $<1000$ ADU taken from sub second exposures, the detector shows good linearity at well within 2\%.

\begin{figure}[t!]
\centering
\includegraphics[trim=0mm 5mm 15mm 10mm, clip, width=84mm]{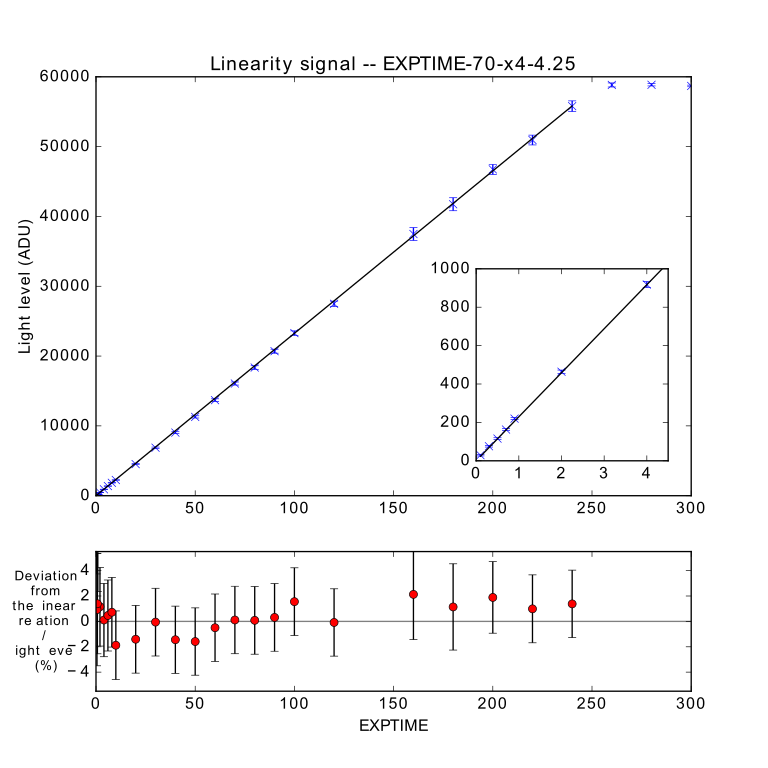}
\caption{Linearity of the CCD as tested in the laboratory. The CCD shows a very good linearity from tens of ADU to the saturation value of $\sim$ 60,000 ADU.}
\end{figure}

\subsection{Shutter Pattern}

To examine at what exposure time the shutter pattern starts to show in the images, we obtained short exposure light images from 0.1 sec to a few sec. Figure 7 shows the light images taken with short exposures. One of the images is the division of the 0.1 sec images by the 6.0 sec image, which shows no significant change in the image pattern from 0.1 to 6.0 sec exposures. Considering the shutter speed is set at 30 msec for the open and close time each, shutter pattern may show up in exposure times well below 0.1 sec. So far, we can conclude that images taken with the exposure time as short as 0.1 sec is not affected by the shutter pattern.

\begin{figure}[t!]
\centering
\includegraphics[trim=25mm 1mm 25mm 1mm, clip, width=84mm]{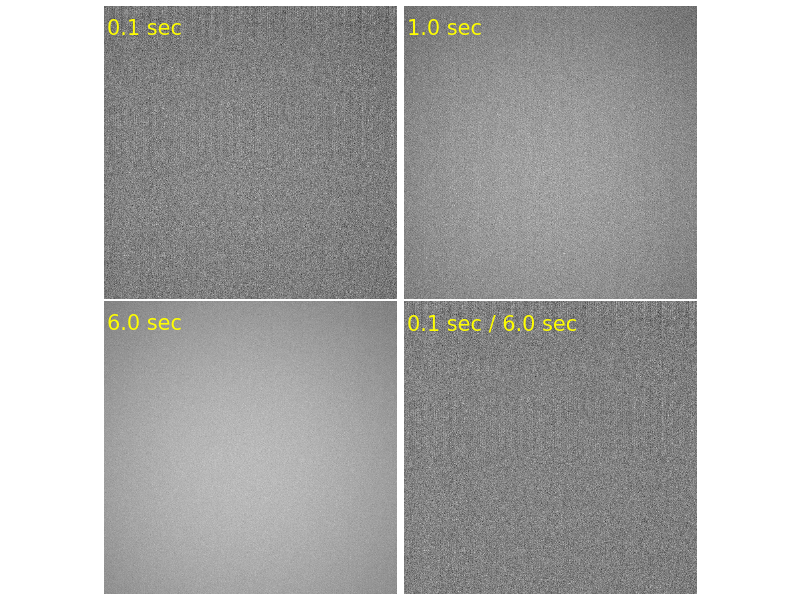}
\caption{Short exposure time light frames that were taken during the laboratory test. The exposure times are indicated in each panel. No significant pattern is visible when 0.1 sec frame is divided by 6.0 sec frame, suggesting that the shutter pattern is negligible even in the 0.1 sec image.}
\end{figure}

\section{Performance\label{sec:performance}}

\subsection{Standard Star Observation, Zero Points,\\ and AB Offsets}

We derived photometric zero points from standard star observations and AB magnitude offset. Standard stars were observed in dark clear nights of seeing condition of FWHM $\sim$ 2.5 arcsec. We observed an A0V star HIP 114918 (V = 9.05 mag) from the \textit{HIPPARCOS} catalog (Perryman et al. 1997) in all the 18 filters on 2016 Oct 31, Nov 02 and 21.
First, we calculated synthetic magnitudes in our filter system using the Vega model spectra of Kurucz (1993) based on equation (3), where F($\lambda$) is the specific flux of Vega model in $\lambda$, and R($\lambda$) is the throughput of the SNUCAM-II system consisting of filter transmission, CCD QE, and the telescope optics throughput. The AB magnitudes of a zero magnitude star were estimated to find the AB offset from Vega magnitude by substituting $F(\lambda) =  10 ^{48.6/2.5} $c$ /\lambda^{2}$ for Vega model flux, where $c$ is the speed of light.
\begin{equation}
{\rm m} =-2.5 \log  \frac{\int F(\lambda)R(\lambda)d\lambda}{\int R(\lambda)d\lambda} + {\rm const}
\end{equation}
\begin{equation}
{\rm M} = {\rm m} + \kappa (X-1) + C
\end{equation}
We measured the magnitudes of the standard star using Source Extractor (Bertin \& Arnouts 1996) auto-mag following equation (4), where $M$ is the apparent magnitude of the star, $m$ is the instrumental magnitude, $\kappa$ is the atmospheric extinction coefficient, X is airmass term ($\sec Z$, where $Z$ is the zenith distance) and $C$ is the zero point at airmass 1. We measured the atmospheric extinction coefficients and the zero point of all the filters and they are presented in Table 3 along with AB offsets. The errors indicate the root mean square (rms) values of three nights data. Note that there was considerable variation in the atmospheric coefficient values in each night, and the presented values should serve only as a rough measure. Figure 8 shows the zero point of each filter at $X = 1$ along with the overall efficiency of the SNUCAM-II, including the telescope throughputs. The most sensitive band filter is $g$ where the overall throughput of the system is the highest and the zero points generally follows the overall throughput of the system. The zero point of the long wavelength region is shallower than that of central wavelength region depending on throughput.

\begin{table}[t!]
\caption{Characteristics of SNUCAM-II }
\centering
\begin{tabular}{lr}
\toprule
Gain                   &   1.15 $\pm$ 0.03 $e^{-}$/ADU             \\
Read noise             &   6.0 $e^{-}$                             \\
Dark current           &   0.2 $e^{-}$/pixel/sec at -70 $^{\circ}$C\\
Pixel scale            &   0.92 arcsec/pixel                       \\
FOV                    &   15.7 arcmin x 15.7 arcmin               \\
Pixel size             &   13 $\mu$m $\times$ 13 $\mu$m              \\
Image area             &   13.312 mm $\times$ 13.312 mm              \\
Cooling temperature    &   -70 $^{\circ}$C                          \\
Digitization precision &   16 Bit                                  \\
Pixel number           &   1024 $\times$ 1024                      \\
A/D readout rate       &   1 MHz                                   \\
Readout time           &   0.88 sec (1 $\times$ 1 binning) \\
Linearity              &   $>> 98$ \% (0 - 60000 ADU)                \\
\bottomrule
\end{tabular}
\end{table}

\begin{table}[t!]
\caption{Photometric calibration parameters from standard star data (Vega system)  \label{tab:snucamtable2}}
\centering
\begin{tabular}{crcc}
\toprule
Filter &   AB   & Atmospheric  & Zero  \\
name   & offset & Extinction   & Point \\
       &        & Coefficient  & (Vega)\\
\midrule
$u$     &  0.864  &  -0.516 $\pm$ 0.142 & 18.698 $\pm$ 0.115 \\
$g$     & -0.106  &  -0.265 $\pm$ 0.078 & 22.178 $\pm$ 0.067 \\
$r$     &  0.156  &  -0.120 $\pm$ 0.026 & 21.679 $\pm$ 0.074 \\
$i$     &  0.383  &  -0.040 $\pm$ 0.104 & 21.195 $\pm$ 0.086 \\
$z$     &  0.525  &  -0.072 $\pm$ 0.079 & 20.419 $\pm$ 0.089 \\
$m425$  & -0.155  &  -0.334 $\pm$ 0.068 & 20.894 $\pm$ 0.080 \\
$m475$  & -0.103  &  -0.158 $\pm$ 0.096 & 20.989 $\pm$ 0.075 \\
$m525$  & -0.039  &  -0.246 $\pm$ 0.012 & 20.894 $\pm$ 0.024 \\
$m575$  &  0.054  &  -0.191 $\pm$ 0.035 & 20.736 $\pm$ 0.021 \\
$m625$  &  0.149  &  -0.140 $\pm$ 0.006 & 20.536 $\pm$ 0.041 \\
$m675$  &  0.267  &  -0.083 $\pm$ 0.069 & 20.325 $\pm$ 0.060 \\
$m725$  &  0.321  &  -0.136 $\pm$ 0.037 & 20.178 $\pm$ 0.067 \\
$m775$  &  0.401  &  -0.044 $\pm$ 0.050 & 19.843 $\pm$ 0.067 \\
$m825$  &  0.479  &  -0.063 $\pm$ 0.078 & 19.602 $\pm$ 0.080 \\
$m875$  &  0.519  &  -0.032 $\pm$ 0.027 & 19.350 $\pm$ 0.048 \\
$m925$  &  0.499  &  -0.121 $\pm$ 0.113 & 18.769 $\pm$ 0.064 \\
$m975$  &  0.546  &  -0.065 $\pm$ 0.097 & 18.106 $\pm$ 0.067 \\
$m1025$ &  0.623  &   0.027 $\pm$ 0.070 & 16.910 $\pm$ 0.052 \\
\bottomrule
\end{tabular}
\end{table}

\subsection{NGC6902 Observation: On-Sky Magnitude Calibration and Magnitude Limits}

To test the on-sky performance of SNUCAM-II, we took images of NGC6902 in all the filters. Figure 9 shows the images of NGC 6902 in all the 18 SNUCAM-II filters taken with 180 sec single exposure. They were taken one after another on the same night with the `filter offset' option in a clear night with seeing FWHM $\sim$ 3 arcsec, where `filter offset' means that the focus changes among various filters were made using a preset amount of the focus shift values between filters. We also derived the zero-point of the images and the image depths from the data. To derive the zero point, we downloaded the $BVgri$ photometric data of APASS (Henden et al. 2016) and $J$, $H$  magnitudes of stars in the field of view of SNUCAM-II  images, and selected stars at $12 < r < 15$ mag as photometric reference. The spectra of 175 stars from the stellar spectral library (Gunn \& Stryker 1983) are fitted to the APASS+2MASS (Skrutskie et al. 2006) photometric data points (up to $J$-band), and the best-fit stellar spectra was chosen with a proper offset to the observed magnitude of the stars in the NGC 6902.
Then, the photometric zero point of each filter was derived for each reference star, by taking the mean of the zero points from the stars. The error in the zero point is estimated as the rms scatter of the zero points from different stars. The 5-$\sigma$ point source detection limits are calculated assuming a seeing condition of 3 arcsec which is slightly worse than the median seeing of LSGT (Im et al. 2015a), the aperture diameter of 3 arcsec with the aperture correction included, and at the airmass as indicated. The zero points and the detection limits are shown in Table 4, and also plotted in Figure 8. The derived values agree with the standard star data result within 0.15 mag. The rms scatters in the zero points indicate that the photometric zero-points can be derived to an accuracy of 0.011 to 0.081 mag from the stars in the science images alone by using the APASS+2MASS data.

\begin{table}[t!]
\caption{Photometric calibration parameters from NGC 6902 data  \label{tab:snucamtable3}}
\centering
\begin{tabular}{ccccc}
\toprule
Filter &  Zero    & Zero  & Airmass  & Limiting  \\
name   &  Point   & Point & $\sec Z$ & Magnitude \\
       & (AB mag) & Error &          & (AB mag)  \\
\midrule
$u$     &  19.646  &  0.081  &  1.024  &  18.39\\
$g$     &  22.073  &  0.023  &  1.155  &  19.91\\
$r$     &  21.747  &  0.016  &  1.135  &  19.51\\
$i$     &  21.450  &  0.024  &  1.112  &  18.96\\
$z$     &  20.867  &  0.012  &  1.024  &  18.20\\
$m425$  &  20.845  &  0.042  &  1.026  &  19.34\\
$m475$  &  20.906  &  0.027  &  1.030  &  19.34\\
$m525$  &  20.870  &  0.025  &  1.038  &  19.29\\
$m575$  &  20.742  &  0.020  &  1.051  &  19.01\\
$m625$  &  20.625  &  0.017  &  1.067  &  18.93\\
$m675$  &  20.504  &  0.018  &  1.087  &  18.84\\
$m725$  &  20.401  &  0.018  &  1.110  &  18.63\\
$m775$  &  20.166  &  0.011  &  1.135  &  18.27\\
$m825$  &  19.952  &  0.015  &  1.168  &  18.06\\
$m875$  &  19.766  &  0.009  &  1.202  &  17.74\\
$m925$  &  19.206  &  0.014  &  1.249  &  17.38\\
$m975$  &  18.548  &  0.014  &  1.295  &  16.74\\
$m1025$ &  17.409  &  0.031  &  1.349  &  15.61\\
\bottomrule
\end{tabular}
\end{table}

\begin{figure}[t!]
\centering
\includegraphics[trim=10mm 2mm 13mm 12mm, clip, width=84mm]{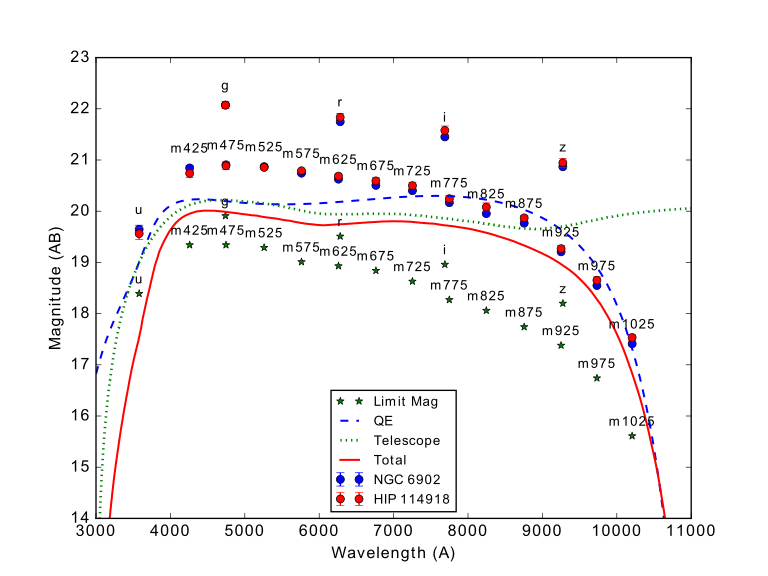}
\caption{Limiting magnitudes (green stars) and magnitude zero points derived from the NGC 6902 field (blue filled circles) and the HIP 114918 data (red filled circles). The blue dashed line is the QE of the CCD camera, the green dotted line is the telescope throughput and the red solid line is the overall throughput of the system. The lines are scaled to match the magnitude scale of the zero point values. The limiting magnitudes assume the exposure time of 180 sec and the 5-$\sigma$ detection of a point source.}
\end{figure}

\section{Scientific Programs\label{sec:Capability}}

By virtue of its many filters and increased sensitivity, LSGT is now more powerful than before for various scientific observations. The 13 medium band filter photometry can provide low resolution spectroscopy of R $\sim$ 15 (Kim et al. 2016). Remote observation of the southern hemisphere has shown its promises for the transient observation of SNe, GRB afterglow (Choi et al. 2016) and gravitational wave sources. LSGT has been one of the main observational facilities of IMSNG (Intensive Monitoring Survey of Nearby Galaxy), which aims to discover newly appearing SNe and detect early shock-heated emission after explosion that can constrain the size of the SN progenitor star (Im et al. 2015b). AGN monitoring study can benefit from medium band photometry that can sample broad emission lines for the reverberation mapping study of AGN black hole masses, and the medium-band reverberation mapping of several AGNs is ongoing. SNUCAM-II on LGST has been used in small research projects for graduate student classes. Examples of the small projects include the SED study of asteroids, SN remnants, variable stars, and stellar clusters.

\begin{figure*}[t!]
\centering
\includegraphics[width=174mm]{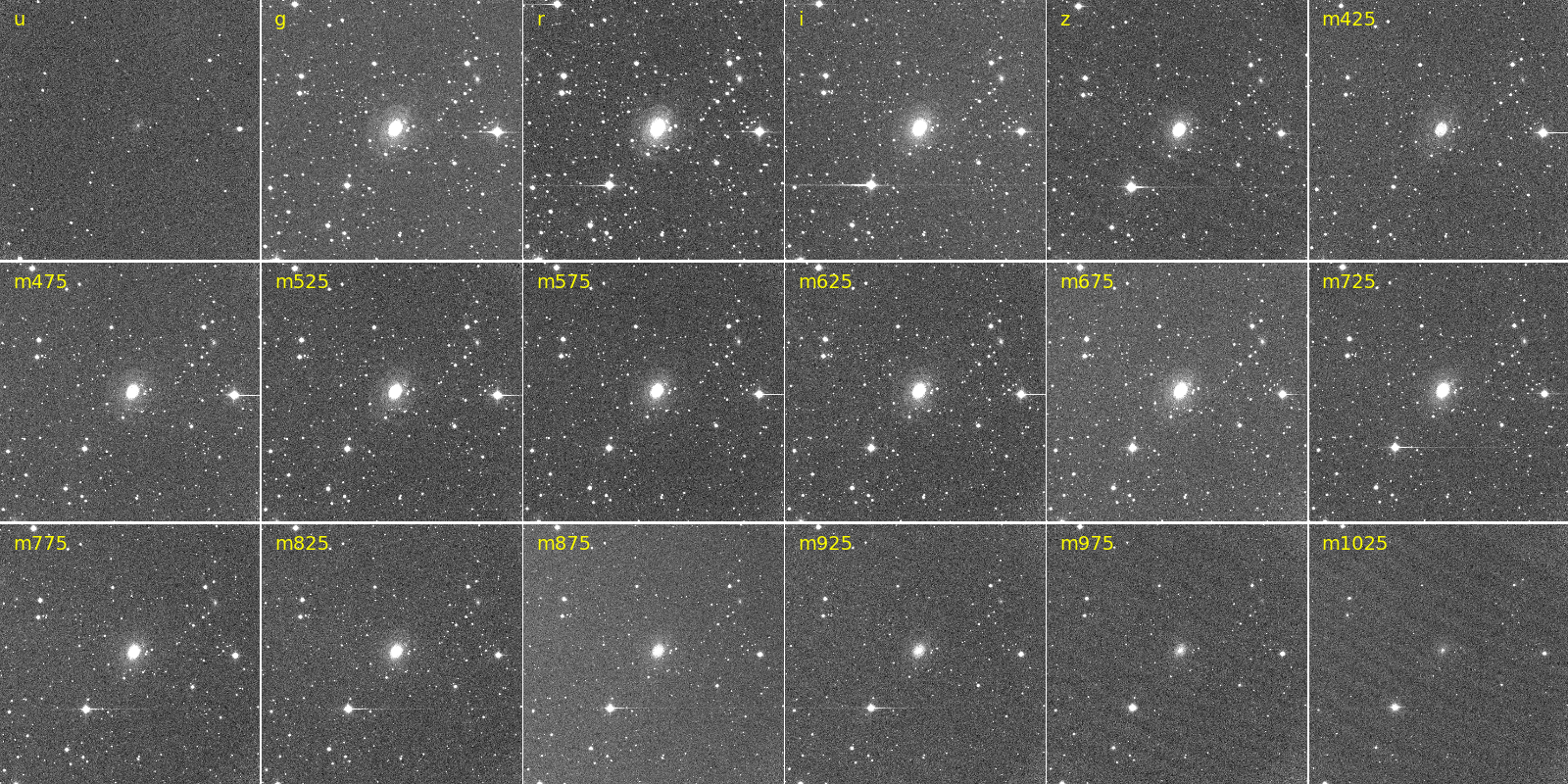}
\caption{Images of NGC 6902 taken with all filters currently installed on SNUCAM-II. Each image was taken with a single 180 sec exposure, and covers the whole field of view of SNUCAM-II (15.7 arcmin $\times$ 15.7 arcmin).}
\end{figure*}

\section{Summary\label{sec:Summary}}

We have presented the characteristics and the performance of the SNUCAM-II system that is installed on LSGT at the Siding Spring Observatory in Australia. This upgraded system is more powerful than the front-illuminated CCD camera systems that have been in use earlier, with QE of $>80$\% from 400--900 nm, factors of a few to tens improvement at short and long wavelengths. SNUCAM-II also boasts 18 filters, $ugriz$ and 13 medium band pass filters from 400 nm to 1100 nm  having 50 nm band width for the characterization of SEDs of many different kinds of sources. Under the adopted operation parameters, the CCD gain is 1.15$\pm$0.03 $e^{-}$/ADU, the readout time is 0.9 sec for the 1k $\times$ 1k frame. The readout noise is 6.0 $e^{-}$, and the dark current is 0.2 $e^{-}$/sec at $-70^{\circ}$C. The SNUCAM-II system shows a good linearity (better than 98\% at the currently measurable limit) ranging from tens of ADU $\sim$ 60000 ADU. The shutter pattern was also examined, and we find no visible shutter pattern in images even with exposure time as short as 0.1 s. Photometric calibration parameters were derived from the analysis of the data of a standard star and reference stars in the vicinity of NGC 6902, showing that SNUCAM-II on LSGT can reach the magnitude limit of $g=$ 19.91 AB mag and $z=$18.20 AB mag at 5-$\sigma$ with 180 sec exposure time for point source detection. With its high sensitivity at short and long wavelengths, the availability of many medium-band filters, and the robotic operation capability, SNUCAM-II on LSGT can be used to perform unique scientific projects such as photometric reverberation mapping of AGNs and intensive monitoring of galaxies to catch the early light curve of SNe.


\acknowledgments
This work was supported by the Creative Initiative program, No. 2017R1A3A3001362, and the grant program No. 2016R1D1A1B03934815 of the National Research Foundation of Korea (NRFK) funded by the Korean government (MSIP). Authors thank to Brad Moore and Pete Lake of iTelescope.net for their help and support. We also thank Won-kee Park for useful discussion.



\end{document}